\documentclass{PoS}
\usepackage{lineno}
\title{Charmonium production in p-Pb collisions with ALICE at the LHC}

\ShortTitle{Charmonium production in p-Pb collisions with ALICE at the LHC}

\author{\speaker{Biswarup Paul}\thanks{on behalf of the ALICE Collaboration}\\
        INFN Torino, Via Pietro Giuria 1, I-10125 Torino, Italy\\
        E-mail: \email{biswarup.paul@cern.ch}}

\abstract{

The suppression of quarkonium production with respect to pp collisions is one of the most distinctive signatures of the formation of quark-gluon plasma (QGP), a hot nuclear medium created in ultrarelativistic heavy-ion collisions. However, the suppression of heavy quarkonium production with respect to pp collisions can also take place in p-A collisions, where QGP is not expected to be created and only cold nuclear matter (CNM) effects, such as nuclear absorption, parton shadowing and parton energy loss in initial and final states occur. The study of p-A collisions is therefore important to disentangle the effects of QGP from the CNM ones, and to provide essential input to understand the nucleus-nucleus collisions.

The ALICE Collaboration at the LHC has studied inclusive J/$\psi$ production, in the dimuon decay channel, at forward rapidity (2.03 $<$ $y_{\rm cms}$ $<$ 3.53) and backward rapidity ($-$4.46 $<$ $y_{\rm cms}$ $<$ $-$2.96) in p-Pb collisions at $\sqrt{s_{\rm NN}} = 8.16$ TeV with the Muon Spectrometer. In this contribution, we present the measurement of the nuclear modification factor of inclusive J/$\psi$ as a function of the centrality and we show the comparison of this result with the one at $\sqrt{s_{\rm NN}} = 5.02$ TeV. 
}

\FullConference{EPS-HEP 2017, European Physical Society conference on High Energy Physics\\
		5-12 July 2017\\
		Venice, Italy}

\begin{document}
\vspace{-3.00mm}
\section{Introduction}
\vspace{-1.500mm}

Cold nuclear matter effects include initial and final state effects. They can either suppress or enhance the quarkonium production. Nuclear shadowing (anti-shadowing) and gluon saturation which affect the partons before the hard scattering are considered as initial state effects. Nuclear absorption and comovers absorption are characterized as final states effects. The shadowing and anti-shadowing describe how the parton distribution function (PDF) of a free nucleon differs from that of a nucleon bound in a nucleus~\cite{shadow1,shadow2}. The gluon saturation describes the coherent interactions of gluons above some saturation scale of $Q^{2}$ in the framework of the colour glass condensate (CGC) effective field theory~\cite{cgc1,cgc2}. The CGC tries to explain the small-$x$ (alternatively high $Q^{2}$) behavior of QCD. The breakup of $Q\overline Q$ pairs due to the inelastic scattering with the nuclear matter surrounding the collision region is called nuclear absorption. The comovers absorption is the suppression of quarkonium in a dense gas system formed by conventional hadrons like pions and kaons. In addition, coherent parton energy loss, in nuclear matter, of the $Q\overline Q$ pair during its evolution is described in Ref~\cite{eloss}. 

The first experimental results on the J/$\psi$ production in p-Pb collisions at $\sqrt{s_{\rm NN}} = 5.02$ TeV were obtained, in 2013, by ALICE~\cite{pPbALICE1,pPbALICE2,5TeVpaper}, ATLAS~\cite{pPbATLAS}, CMS~\cite{pPbCMS} and LHCb~\cite{pPbLHCb} at the LHC. The new results on the nuclear modification of J/$\psi$ vs centrality in p-Pb collisions at $\sqrt{s_{\rm NN}} = 8.16$ TeV presented in this note are complementary to the rapidity and transverse momentum dependence of J/$\psi$ production at the same energy~\cite{pPbALICE8TeV}, both at forward and backward rapidities.
\vspace{-1.500mm}
\section{Experimental detectors and data analysis}
\vspace{-1.500mm}
The ALICE Collaboration has studied inclusive J/$\psi$ production in p-Pb collisions at $\sqrt{s_{\rm NN}} = \mbox{5.02 TeV}$ and 8.16 TeV. In the analysis discussed in this proceedings, the J/$\psi$ is studied in its $\mu^{+}\mu^{-}$ decay channel. Due to the beam-energy asymmetry during the p-Pb data-taking, the nucleon-nucleon center-of-mass (cms) system is shifted in rapidity with respect to the laboratory frame by $\Delta y$ = 0.465 towards the proton beam direction. The data have been taken with two beam configurations, obtained by inverting the directions of the p and Pb beams. This results in the following two rapidity ranges covered with the Muon Spectrometer: forward (2.03 $<$ $y_{\rm cms}$ $<$ 3.53) and backward ($-$ 4.46 $<$ $y_{\rm cms}$ $<$ $-$ 2.96). Muons are identified and tracked in the Muon Spectrometer, which covers the pseudorapidity range $-4<\eta<-2.5$~\cite{alice}. The Silicon Pixel Detector (SPD) which is used for vertexing, consists of two cylindrical layers covering $|\eta|$ $<$ 2.0 and $|\eta|$ $<$ 1.4 for the inner and outer layers, respectively. The V0 consists of two arrays of 32 scintillators (V0A and V0C), placed at $2.8 \leq \eta \leq 5.1$ (V0A) and $-3.7 \leq \eta \leq -1.7$ (V0C). These detectors help to remove the beam induced background and a coincidence of their signals provides the minimun bias (MB) trigger. In this analysis a dimuon trigger is used, based on the coincidence of the MB trigger and two opposite-sign muon tracks in the muon trigger chambers. Two sets of Zero Degree Calorimeters (ZDCs), each including a neutron (ZN) and a proton (ZP) calorimeter, positioned symmetrically at $\pm$ 112.5 m from the interaction point, are used for the centrality estimation. Events where two or more interactions occurring in the same colliding bunch (in-bunch pile-up) or during the readout time of the SPD (out-of-bunch pile-up) are removed using the information from SPD and V0.

In p-Pb collisions, the centrality selection is defined by a selected range of energy deposited by neutrons in the Pb-remnant side of ZN using the hybrid method described in~\cite{centrality}. In this method, the determination of the average number of binary nucleon collisions $\langle N^{\rm mult}_{\rm coll} \rangle$ relies on the assumption that the charged-particle multiplicity measured at mid-rapidity is proportional to the number of participant nucleons ($\langle N_{\rm part} \rangle$). $\langle N_{\rm part} \rangle$ is calculated from the Glauber model~\cite{glauber} which is generally used to calculate geometrical quantities of nuclear collisions. Other assumptions to derive $\langle N_{\rm coll} \rangle$, which are discussed in~\cite{centrality}, are used in order to determine the associated systematic uncertainty. The centrality classes 0-2\% and 90-100\% are excluded due to the possible contamination from residual pile-up events or centrality calibration issues.

In order to improve the purity of the muon tracks, the following selection criteria were applied: (1) tracks reconstructed in the Muon Spectrometer tracking chambers should match the track reconstructed in the trigger system, (2) both muon tracks are in the pseudo-rapidity range $-$4 $<$ $\eta$ $<$ $-$2.5, (3) transverse radius coordinate of the tracks at the end of the hadron absorber (the longitudinal position of the absorber from the interation point (IP) is $-$5.0 $<$ $z$ $<$ $-$0.9 m) is in the range 17.6 $<$ $R_{\rm{abs}}$ $<$ 89.5 cm, (4) dimuon rapidity is in the range  2.5 $<$ $y$ $<$ 4 and (5) dimuon $p_{\rm T}$ is in the range $p_{\rm T}$ $<$ 20 GeV/$\it{c}$. These cuts eliminate mainly tracks hitting the edges of the spectrometer acceptance or crossing the thicker part of the absorber.

J/$\psi$ and $\psi$(2S) yields are extracted by fitting the dimuon invariant mass distributions with a superposition of signals and background shapes. For the signal, pseudo-Gaussian or Crystal Ball functions with asymmetric tails on both sides of the resonance peak are used, while for the background a Gaussian with a mass-dependent width or polynomial $\times$ exponential function are adopted. More details on the analysis are reported in Ref.~\cite{pPbALICE8TeV,pPbALICE8TeV2}.
\begin{figure}[ht]
\includegraphics[scale=0.38]{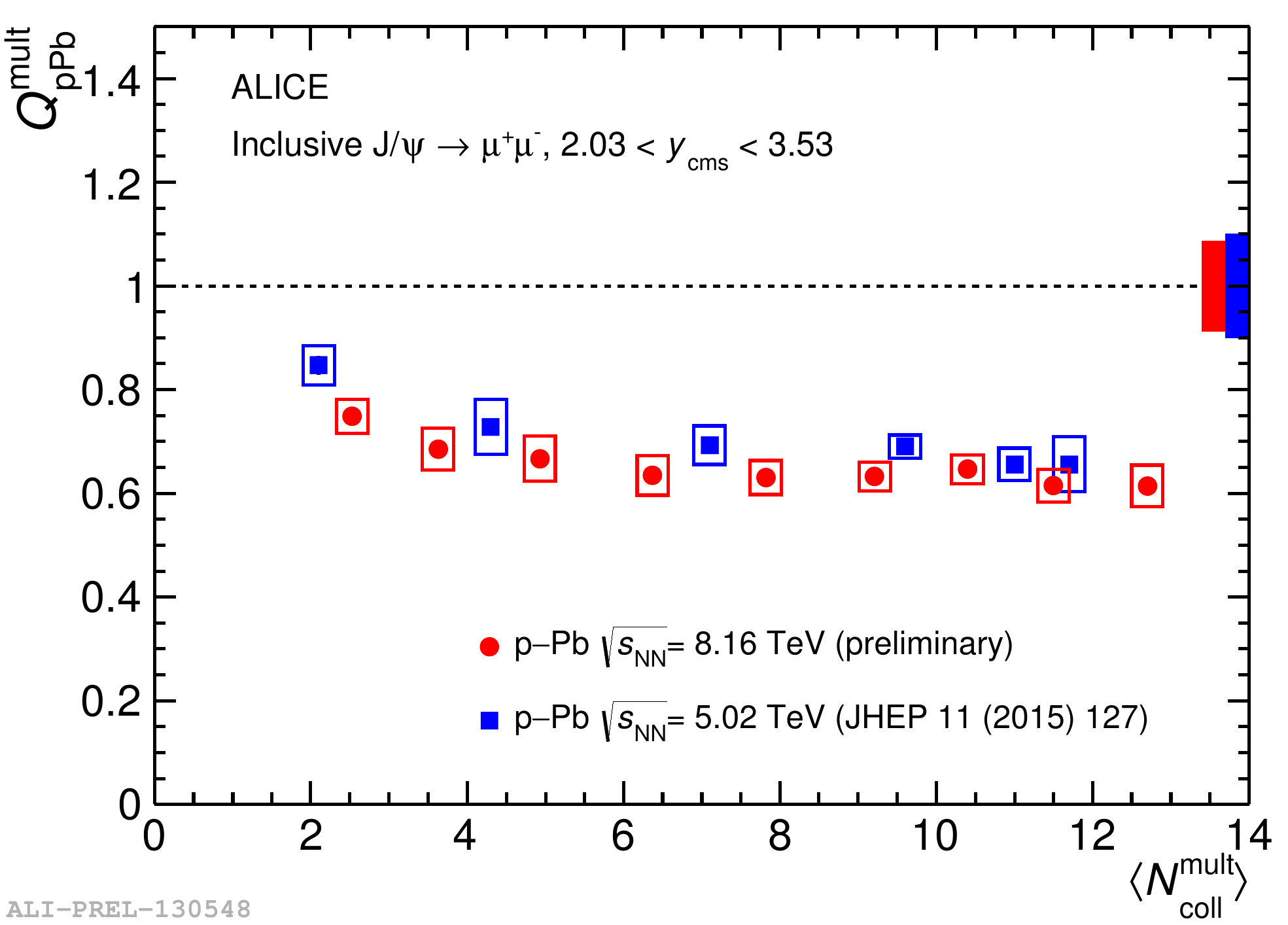}
\includegraphics[scale=0.38]{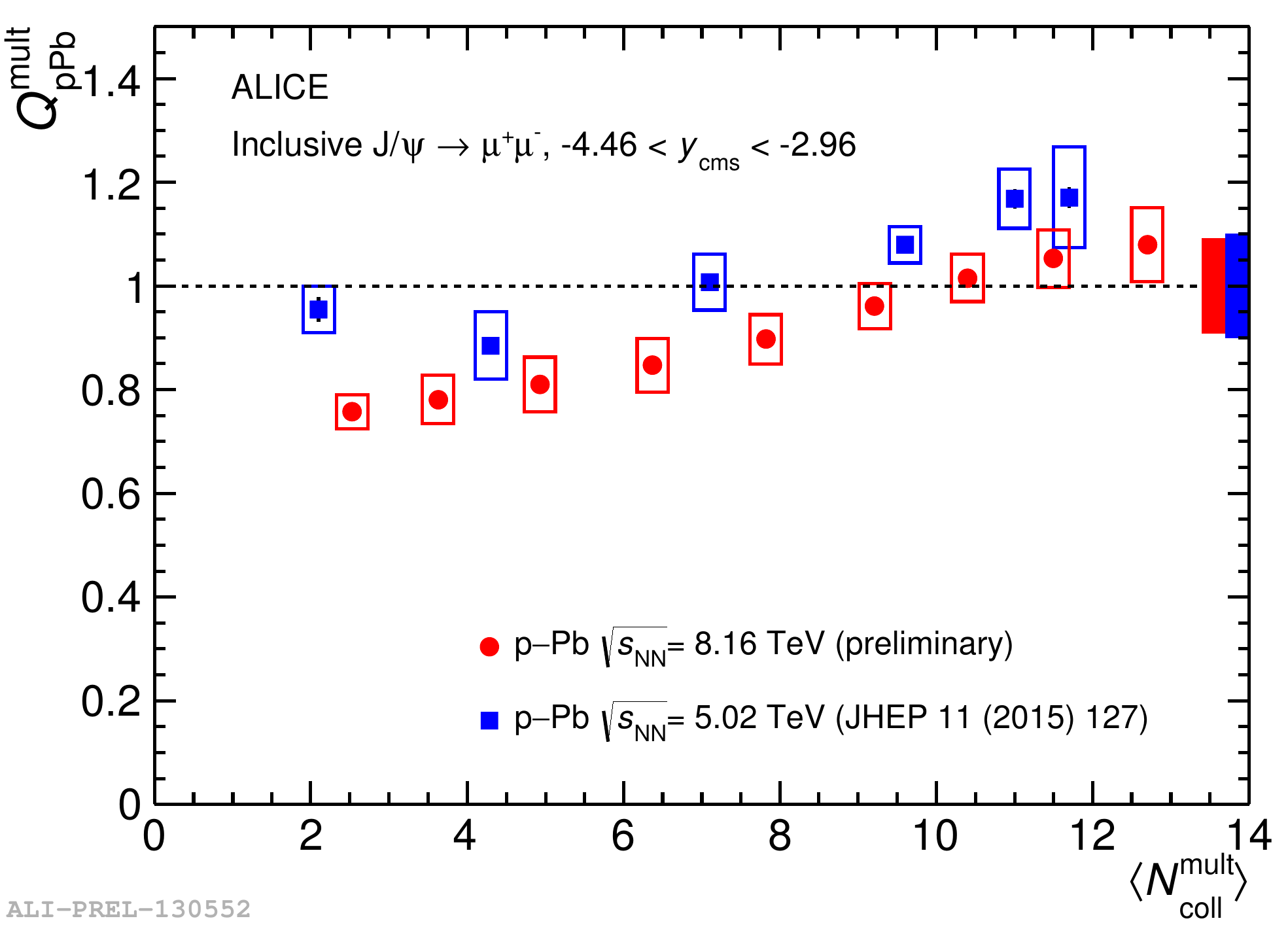}
\caption{\label{QpPb8TeV}Nuclear modification factor of J/$\psi$ as a function of centrality at forward (left) and backward (right) rapidity at $\sqrt{s_{\rm{NN}}}$ = 8.16 TeV compared to those obtained at $\sqrt{s_{\rm{NN}}}$ = 5.02 TeV in the same centrality classes, except for the most peripheral bin, which corresponds to 80-90\% for the results at $\sqrt{s_{\rm{NN}}}$ = 8.16 TeV and 80-100\% for those at $\sqrt{s_{\rm{NN}}}$ = 5.02 TeV. The statistical uncertainties are shown as lines and the systematic uncertainties are shown as boxes around the points, while the box around unity corresponds to the global uncertainty.}
\end{figure}
\vspace{-3.00mm}
\section{Results}
\vspace{-1.500mm}
The nuclear modification factor is defined as the ratio of the quarkonium production yield in p-Pb collisions to that in pp collisions scaled with $\langle N^{\rm mult}_{\rm coll} \rangle$. In ALICE for the centrality-dependent studies in p-Pb collisions it is referred to as $Q^{\rm mult}_{\rm pPb}$ due to the possible bias in the determination of centrality. The $Q^{\rm mult}_{\rm pPb}$ of inclusive J/$\psi$ as a function of ZN centrality classes in p-Pb collisions at $\sqrt{s_{\rm NN}} = 8.16$ TeV is shown in Fig.~\ref{QpPb8TeV} and the results have been compared with the previous ALICE results at $\sqrt{s_{\rm NN}} = 5.02$ TeV~\cite{pPbALICE8TeV2}. The high luminosity data collected at $\sqrt{s_{\rm NN}} = 8.16$ TeV allows us to study $Q^{\rm mult}_{\rm pPb}$ in narrow centrality classes, increasing the precision of our measurement w.r.t $\sqrt{s_{\rm NN}} = 5.02$ TeV. The J/$\psi$ shows a suppression which increases slightly from the peripheral bin (80-100\% at $\sqrt{s_{\rm NN}} = 5.02$ TeV and 80-90\% at $\sqrt{s_{\rm NN}} = 8.16$ TeV) to the most central one (2-10\%) at forward rapidity while at backward rapidity, $Q^{\rm mult}_{\rm pPb}$ has an opposite trend, showing a suppression in the most peripheral bin with a increasing pattern towards the most central events. A similar pattern is observed at the two center-of-mass energies which confirms the consistency already observed in the $p_{\rm T}$ and rapidity dependence of $R_{\rm pPb}$~\cite{pPbALICE8TeV}. The results at the two energies are compatible within the uncertainties of the measurement.
\vspace{-2.00mm}
\section{Conclusions}
\vspace{-1.500mm}
The ALICE Collaboration has measured the centrality dependence of the inclusive J/$\psi$ production in p-Pb collisions at $\sqrt{s_{\rm NN}} = 8.16$ TeV with high  precision and the results are compatible with the same measurement at $\sqrt{s_{\rm NN}} = 5.02$ TeV. The results show a suppression which slightly increases towards central collisions at forward rapidity while at backward rapidity a different trend is observed, with the nuclear modification factor increasing from peripheral to central collisions.


\vspace{-1.00mm}
\begin{thebibliography}{99}
\vspace{-2.00mm}
\bibitem{shadow1} M. Hirai, S. Kumano, and T. H. Naga, {\em Phys. Rev. C {\bf 76} (2007) 065207}.
\bibitem{shadow2} K. J. Eskola, H. Paukkunen, and C. A. Salgado, {\em JHEP {\bf 04} (2009) 065}. 
\bibitem{cgc1} F. Gelis, E. Iancu, J. Jalilian-Marian, and R. Venugopalan, {\em Ann. Rev. Nucl. Part. Sci. {\bf 60} (2010) 463}.
\bibitem{cgc2} H. Fujii, F. Gelis, and R. Venugopalan, {\em Nucl. Phys. A {\bf 780} (2006) 146}.
\bibitem{eloss} F. Arleo and S. Peigne, {\em JHEP {\bf 03} (2013) 122}.
\bibitem{pPbALICE1} {\bfseries ALICE} Collaboration, {\em JHEP {\bf 02} (2014) 073}.
\bibitem{pPbALICE2} {\bfseries ALICE} Collaboration, {\em JHEP {\bf 06} (2015) 055}.
\bibitem{5TeVpaper} {\bfseries ALICE} Collaboration, {\em JHEP {\bf 11} (2015) 127}.
\bibitem{pPbATLAS} {\bfseries ATLAS} Collaboration, {\em Phys. Rev. C {\bf 92} (2015) 034904}.
\bibitem{pPbCMS} {\bfseries CMS} Collaboration, {\em Eur. Phys. J. C {\bf 77} (2017) 269}.
\bibitem{pPbLHCb} {\bfseries LHCb} Collaboration, {\em JHEP {\bf 02} (2014) 072}.
\bibitem{pPbALICE8TeV} {\bfseries ALICE} Collaboration, ALICE-PUBLIC-2017-001, https://cds.cern.ch/record/2244670
\bibitem{alice} {\bfseries ALICE} Collaboration, {\em JINST} {\bf 3}, (2008) S08002.
\bibitem{centrality} {\bfseries ALICE} Collaboration, {\em Phys. Rev. C {\bf 91} (2015) 064905}.
\bibitem{glauber} M. L. Miller, K. Reygers, S. J. Sanders and P. Steinberg, {\em Ann. Rev. Nucl. Part. Sci. {\bf 57} (2007) 205}.
\bibitem{pPbALICE8TeV2} {\bfseries ALICE} Collaboration, ALICE-PUBLIC-2017-007, https://cds.cern.ch/record/2272151
\end{thebibliography}
\end{document}